# Neutron background measurement for rare event search experiments in the YangYang Underground Laboratory

Young Soo Yoon, Jungho Kim, Hyeonseo Park[*]

*Korea Research Institute of Standards and Science, Daejeon, 34113 Republic of Korea*

**Abstract**

Several experiments have been conducted in the YangYang Underground Laboratory in the Republic of Korea such as the search for dark matter and the search for neutrinoless double beta decay, which require an extremely low background event rate due to the detector system and the environment. In underground experiments, neutrons have been identified as one of the background sources. The neutron flux in the experimental site needs to be determined to design a proper shielding system and for precise background estimation. We measured the neutron spectrum with a Bonner sphere spectrometer, with Helium-3 ($^3$He) proportional counters. The neutron flux at the underground laboratory was so low that the radioactive decays from the radioisotopes contained in the detector created a significant background interference to the neutron measurement. Using Monte Carlo simulations, the intrinsic $\alpha$ background distribution due to the radioactive isotopes in the detector materials, was estimated. The neutron count rate of each Bonner sphere was measured from the pulse height spectrum of the $^3$He proportional counter, after subtracting the $\alpha$ particle background. The neutron flux and the energy spectrum were determined using the unfolding technique. The total neutron flux measured was $(4.46 \pm 0.66) \times 10^{-5}$ cm$^{-2}$ s$^{-1}$, and the thermal and fast neutron flux (in the range 1 to 10 MeV) were $(1.44 \pm 0.15) \times 10^{-5}$ cm$^{-2}$ s$^{-1}$ and $(0.71 \pm 0.10) \times 10^{-5}$ cm$^{-2}$ s$^{-1}$, respectively.

[*]Corresponding author: hyeonseo@kriss.re.kr

1. **Introduction**

Rare event search experiments, such as the search for dark matter or neutrinoless double beta decay, require an extremely low background event rate due to the detector system and the environment. Therefore, most experiments are conducted in underground laboratories to reduce the effect of cosmic rays. In underground experiments, neutrons are known as one of the background sources. In dark matter experiments that look for weakly interacting massive particles (WIMPs), nuclear recoils produced by neutrons are indistinguishable from those produced by WIMPs. This forms a critical background source for the experiments. In experiments that search for neutrinoless double beta decay, $\gamma$-rays with energies of several MeV can be emitted from neutron inelastic scattering (n, n'$\gamma$) or capture (n, $\gamma$). These emitted $\gamma$-rays cover the region of interest of the search for neutrinoless double beta decay.

The neutrons present in underground laboratories come from cosmic-ray muon interactions, ($\alpha$, n) reactions induced by $\alpha$-particles from the radioactive isotopes in rock, and the spontaneous fission of Uranium-238($^{238}$U) [1]. In an underground laboratory at a depth of 2000 meter-water-equivalent (m.w.e.), the neutron flux induced by cosmic muons is approximately two orders of magnitude lower than that induced by nuclear reactions [1].

In Korea, the YangYang Underground Laboratory has been used for the Korea Invisible Mass Search (KIMS) experiments [2] since 2003. The laboratory is situated in one of the tunnel spaces (A6) in the YangYang Pumped Storage Power Station of the Korea Hydro & Nuclear Power Co., where the rock overburden is approximately 700 m (~2000 m.w.e.). The neutron flux in the KIMS experimental site has been measured and reported [3]. In another tunnel space (A5) in the YangYang Pumped Storage Power station, several hundred meters away and at a similar depth, two laboratories were built in 2014. They were intended for a dark matter search experiment called COSINE [4], and a search for neutrinoless double beta decay, the Advanced Mo-based Rare process Experiment (AMoRE) [5, 6]. Several experiments have been conducted there since, including simulation studies and background measurements for the AMoRE [7] to investigate background sources. As a result, neutrons were identified as one of the background sources. The neutron flux in the experimental site needs to be determined to design a proper shielding system and for precise background estimation.

Various measurement techniques have been used for neutron measurements in underground laboratories, such as liquid scintillators [8, 9, 10] for fast neutrons, proportional counters [11, 12] for thermal neutrons, and proportional counters with moderators for wide range neutrons [3, 13, 14]. In this study, we measured the neutron spectrum in the A5 tunnel at YangYang. The measurements were carried out using Bonner sphere spectrometers (BSSs) with Helium-3 ($^3$He) proportional counters [15].

## 2. Experiment

### 2.1. Bonner sphere spectrometer

The neutron spectrum and flux in the underground laboratory were measured using a BSS system. The BSS is an instrument that is widely used to evaluate neutron energy and its flux [3, 15, 16, 17]. A BSS consists of neutron moderating material and a thermal neutron detector. The moderator is spherically shaped to get the isotropic response and the thermal neutron detector is located at the center of the moderator. After entering the detector, the fast neutrons are moderated after several collisions inside the moderator material, and when they reach the thermal neutron detector they make a signal. The size of the moderator determines the neutron response according to its energy, i.e. a large moderator gives a high response to high energy neutrons and a smaller moderator gives a high response to low energy neutrons.

In this study, we used nine spherical thermal neutron detectors (SP9 $^3$He proportional counters filled with $^3$He at 2 atm) coupled to nine Bonner spheres (BSs) with diameters ranging from 7.6 cm to 25.4 cm, and one bare thermal neutron detector.

This set of BSSs was designed by Physikalisch-Technische Bundesanstalt [18] and manufactured by Centronic [19]. Table 1 shows the BSSs used in the measurement and Fig. 1 shows the response function of the BSSs as a function of the neutron energy. The response was calculated using the Monte Carlo N-Particle Extended 2.70 code, called NCNPX, and calibrated using a Californium ($^{252}$Cf) neutron source at the Korea Research Institute of Standards and Science (KRISS).

This configuration of the BSS system can be used for neutrons with energy less than 20 MeV. For higher energy neutrons, the bigger size BS or BSs having a metal shell (multi-shell BS) should be used. The previous measurements using multi-shell BSs at the YangYang Underground Laboratory showed that neutrons over 20 MeV accounted for approximately 1% of all neutrons [3]. Because of the extremely low response of the BSs used for this measurement and the low rate of high energy neutrons, we restricted the energy range to a maximum of 20 MeV.

Table 1: BSS sets used for this study.

| Bonner Sphere Id | Diameter of moderator (cm) | Live time (days) |
|---|---|---|
| KB010 | No moderator | 67.5 |
| KB030 | 7.62 | 67.3 |
| KB035 | 8.89 | 93.7* |
| KB040 | 10.2 | 67.4 |
| KB045 | 11.4 | 93.7* |
| KB050 | 12.7 | 67.4 |
| KB060 | 15.2 | 67.1 |
| KB070 | 17.8 | 67.3 |
| KB080 | 20.3 | 67.6 |
| KB100 | 25.4 | 67.5 |

* A data acquisition laptop with KB035 and KB045 operated for the whole period, while the 2nd laptop with the rest of the detectors operated for 68 days.

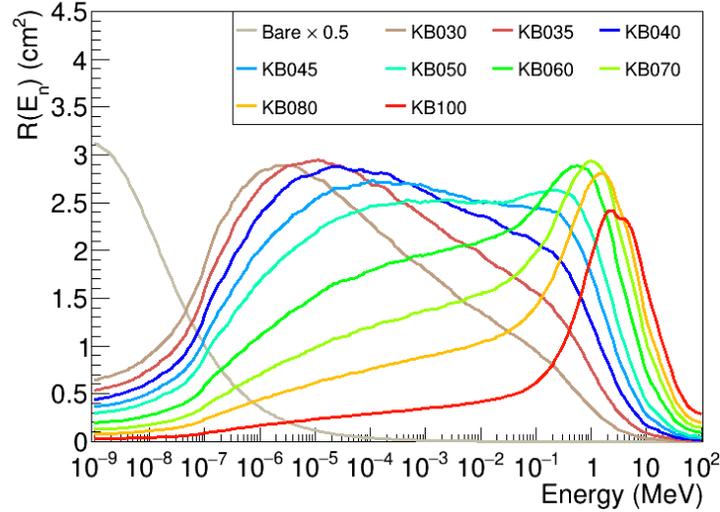

Figure 1: Calculated response function of BS of KRISS.

*2.2. Measurement in the underground laboratory*

The measurements were conducted in the underground laboratory in the A5 tunnel at YangYang. The BSS system was installed approximately three meters from the main detector set-up of the neutrinoless double beta decay search experiment, which included copper shields, a huge mass of lead shield, polyethylene shield, plastic scintillators, and an iron structure frame. Figure 2 shows the configuration of the detector set-up for the measurement. Because of the limited space, BSs and their signal processing modules were installed at three levels. Each BS had its own thermal neutron detector (SP9) and pre-amplifier/amplifier/HV modules. The signals from 10 channels were collected by four multichannel buffers (MCBs, EG&G Ortec model 919E and ASPEC-927). The dynamic range of each MCB was 10 V with 1024 channels. The pulse height spectrum of each BS channel was collected and saved once every hour. In this way, the data loss due to external conditions was minimized. The status of the detector system was monitored remotely via the internet at the ground laboratory. The measurements were taken over a 94-day period. The live time of each BS is summarized in Table 1. The pulse height spectrum of each BS was obtained after summing all one-hour spectra.

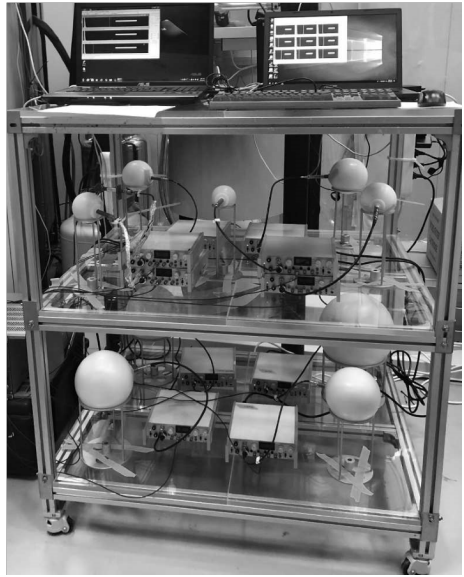

Figure 2:   BSS system installed in A5 laboratory, YangYang

## 3.  Analysis

*3.1.  Neutron detection*

The $^3$He proportional counter detected thermal neutrons by measuring the products of the following reaction:

$$^3He + n \rightarrow {}^3H + p \text{ (Q = 0.764 MeV)}. \quad (1)$$

Fully ionized products of the reaction in the counter were observed as peaking at 764 keV in the pulse height spectrum. Figure 3 shows a typical pulse height spectrum of an SP9 for thermal neutrons. The spectrum consists of a full energy peak and a flat distribution due to the wall effect. The gamma or noise contribution can be seen in the region with a low pulse height, which is clearly separated from the thermal neutrons in the source measurement. The thermal neutron range for each SP9 was determined from the middle of the valley between the gamma and neutron range to the maximum signal of the full peak area, as shown in Fig 3. The pulse height spectra for the thermal neutrons were obtained for all SP9s before the underground measurement, using an Americium-241-Beryllium ($^{241}$Am-Be) neutron source. These pulse-height spectra were used as the reference spectra during the analysis for each BS. The pulse height was converted to the energy scale using the full peak position of the thermal neutrons corresponding to 764 keV.

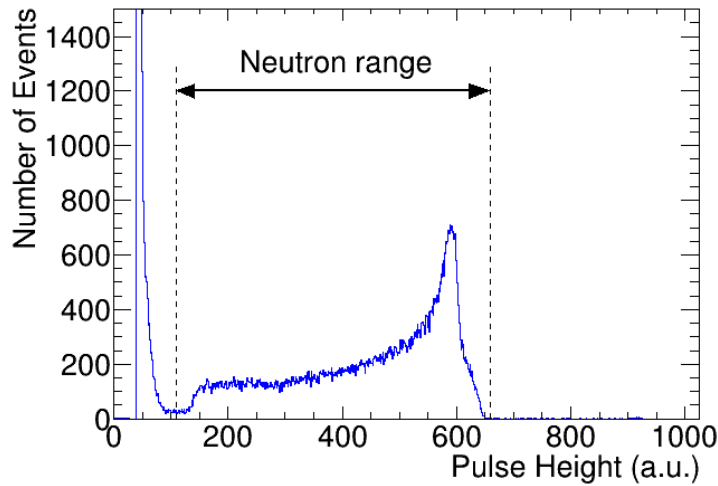

Figure 3: Typical pulse height spectrum of an SP9 for thermal neutrons.

*3.2. Pulse-height spectra of neutron detectors*

Figure 4 shows the actual pulse-height distribution of an SP9 obtained with the underground measurements (blue histogram). The reference spectrum obtained in the ground laboratory using the $^{241}$Am-Be neutron source is overlaid after scaling down for comparison (red histogram). Both spectra can be separated into three components: gamma/noise signals in the low pulse-height region, thermal neutron signals in the middle region, around 150–650 channels (labeled as the neutron region), and the high pulse-height backgrounds, above 650 channels, which are not visible in the red spectrum. The red spectrum shows a clear separation between the neutrons and the gamma rays. However, in the blue spectrum, because of the long exposure needed for the extremely low neutron rate, the tail of the gamma signals from the environment persisted until approximately channel 200. The separation between the neutrons and gamma rays is not clear. The high pulse-height backgrounds are shown only in the blue spectrum, which is assumed to be the intrinsic backgrounds of the SP9. Because they appeared following long exposure and the intrinsic background level is quite low, it is not visible in the reference spectrum, which was measured for a few hours only. In the underground measurements the intrinsic background level is comparable to the neutron rate, and hence, is significant. The narrow peak above 900 channels in the blue spectrum indicates an overflow due to the limited dynamic range. The intrinsic background signals due to decays of radioisotope contamination, such as $^{238}$U and $^{232}$Th, in the detector materials.

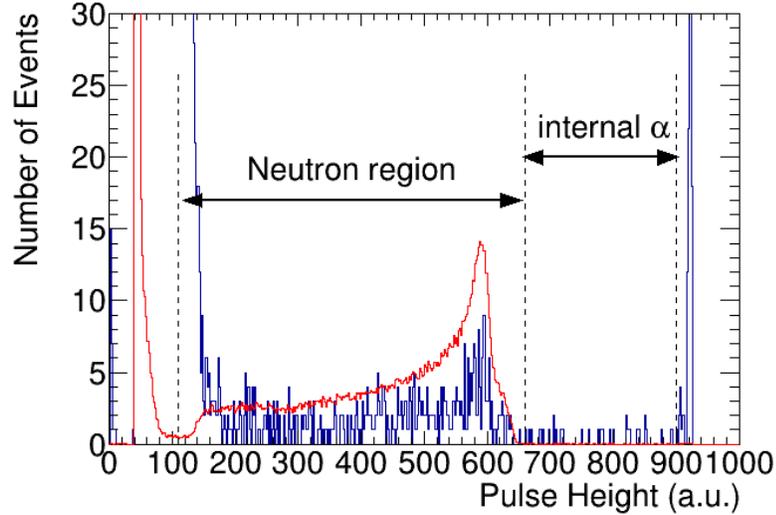

Figure 4: Pulse-height spectrum measured in the underground laboratory (blue) and a reference spectrum measured on the surface (red)

3.3. *Intrinsic background simulations*

The intrinsic background consists mostly of $\alpha$ particles originating from decays of radioactive isotopes [12, 14, 20] inside the detector materials, especially the inner surface of the SP9s. When radioactive isotopes, such as $^{238}$U, $^{232}$Th, and their progeny decay near the inner surface of the detector shell the daughter nuclei can escape into the $^{3}$He-gas-filled internal space of the SP9. The $\alpha$ particles escaping from the surface of the detector have a continuous energy spectrum up to the Q-value of the decay process that extends throughout the neutron region. To determine the deposited energy of the $\alpha$ particles in the $^{3}$He gas, the decay of the radioactive isotopes in the detector was simulated using the GEANT4 toolkit with the built-in G4RadioactiveDecay database [21, 22]. It was assumed that the decay chains of $^{238}$U and $^{232}$Th are in secular equilibrium.

The expected deposited energy distribution in the $^3$He gas in an SP9 due to the decays of $^{238}$U, $^{232}$Th, and their progeny is shown in Fig. 5. The escaping $\alpha$ particles from the detector's inner surface have a flat energy distribution below their Q-value. The various Q-value energies of the $\alpha$ decay channels of $^{238}$U, $^{232}$Th, and their progeny cause step structures in the spectra over a range of 4 MeV, which vary according to the relative concentrations of the $^{238}$U and $^{232}$Th, as shown in Fig. 5 (left plot).

The simulation was performed with relative contamination ratios of 1:1, 1:0.2, and 1:5 for the $^{238}$U and $^{232}$Th. The simulation results show that the spectral shape of the deposited energy of internal $\alpha$ particles does not depend on the concentration of the $^{238}$U and $^{232}$Th in the region of interest, as shown in Fig. 5 (right plot).

$\alpha$ particles from a detector shell with a depth of a few $\mu$m lose most of their energy inside the shell. They enter the sensitive area of the detector with a kinetic energy of less than several hundred keV. The effect of these low energy $\alpha$ particles on the pulse height spectrum requires further investigation. In addition, the ionization loss of the $\alpha$ particles in the $^3$He gas relative to the pulse height of the detector needs to be studied quantitatively to use the simulation results. More studies are ongoing, and in this current study, a constant loss was assumed as the background shape. The count rate difference between a constant function and the fit function was less than 4% in each detector and was included in the systematic uncertainties.

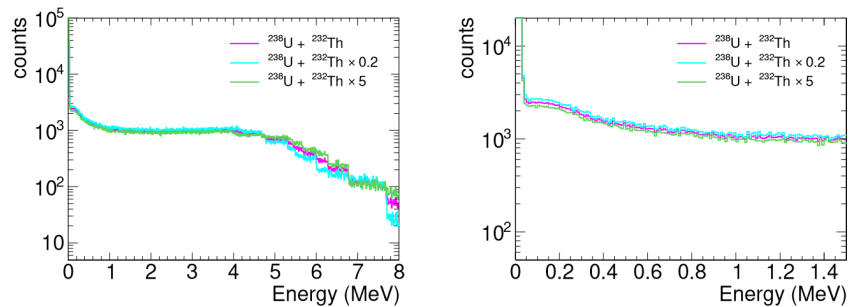

Figure 5: Simulated deposited energy spectrum by $\alpha$ particles from full decay chain simulation of $^{238}$U and $^{232}$Th in the energy ranges up to 8 MeV (left) and 1.5 MeV (right).

*3.4. Neutron count rate*

The neutron count rate was estimated with a binned maximum likelihood fit. The pulse height spectrum for thermal neutrons obtained with the $^{241}$Am-Be source and moderator was used as a reference spectrum, and the background spectrum due to the $\alpha$ particles was assumed to be constant in the fit range. The energy calibration of the pulse-height spectrum of the SP9 was performed using the 764 keV peak. The expected count in the $i^{th}$ bin, $\mu_i$, is defined as $\mu_i = S_{src,i} + S_{bkg,i}$, where $S_{src,i}$ is the neutron count in the $i^{th}$ bin and $S_{bkg,i}$ is the background count in the $i^{th}$ bin. The maximum likelihood, which is equivalent to the minimum of −log-likelihood, $-ln\ L$, is defined with a Poisson distribution as follows:

$$-ln\ L = \sum_i -ln\left(\frac{\mu_i^{n_i} e^{-\mu_i}}{n_i!}\right). \tag{2}$$

$-ln\ L$, was calculated using data in the neutron range and $\alpha$ background range. The optimized fitting range was determined by excluding low-energy gamma events and saturated $\alpha$ events near 1.2 MeV. The range was found to be from 50% (0.38 MeV) to 150% (1.1 MeV) of the 764 keV peak position for each SP9. Figure 6 shows a typical fit result with a measured pulse height spectrum (black), a likelihood fit result (red), and the fit components consisting of neutrons (blue) and background (pink). The fitted spectrum is close to the measurement result above ∼300 keV.

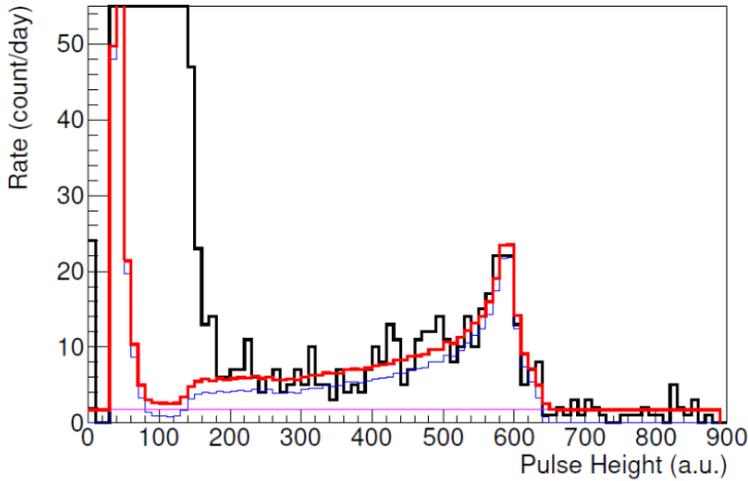

Figure 6: Likelihood fit result (red) of measured pulse height spectrum of SP9 (black) with neutrons (blue) and background (pink).

The neutron count rate was obtained from the fit results and data acquisition live time for each BS. Figure 7 shows the neutron count rate per day for each BS. The uncertainty was estimated from the fit using a range of $-\ln L + 0.5$ on each parameter space, which is equivalent to the statistical uncertainty of the measurements and background subtraction.

The neutron count rate was validated by an alternative method, called a counting method. We integrated the neutron count rate in the likelihood fit region (0.38 MeV to 0.85 MeV) after subtracting the intrinsic $\alpha$ background. The intrinsic background rate per bin was calculated using the event rate in the $\alpha$ region (0.9 MeV to 1.2 MeV). The total neutron count rate can be determined by multiplying the neutron count rate in the likelihood fit region by the area ratio of the full neutron region to the likelihood fit region pre-determined with the reference spectrum. The estimated neutron count rate of the counting method in each detector was consistent with that of the likelihood fit. The difference in neutron count rates between the likelihood fit and the counting method for the BSs was approximately one standard deviation.

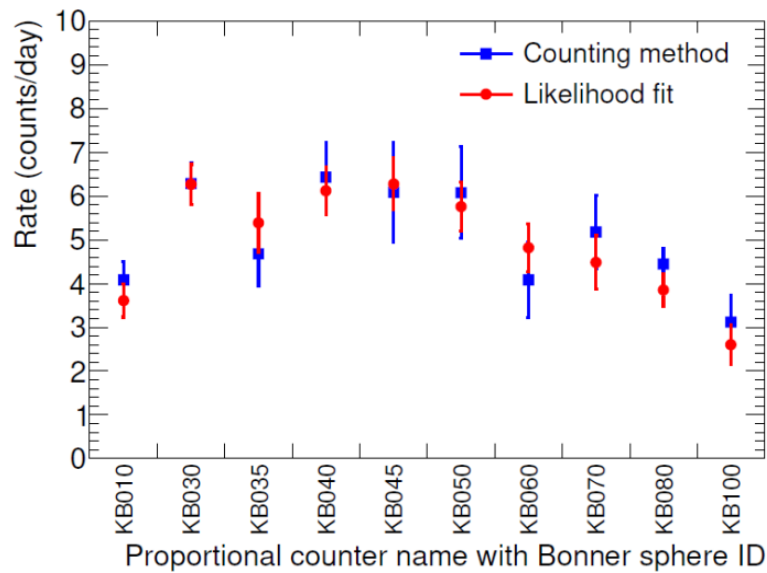

Figure 7: The daily neutron count rate of each SP9 detector with BSs, using the counting method (filled square) and the likelihood fit (filled circles).

3.5. *Neutron spectrum unfolding*

The neutron differential flux, (neutrons per unit area and time), were estimated through an unfolding procedure using the MAXED code. The MAXED code is a process that uses the maximum entropy principle [23].

In some cases, the results of the unfolding could be biased by the shape of the input neutron spectrum of the unfolding procedure. To avoid input spectrum bias, several input spectra were tested with the unfolding process. The previously measured neutron spectrum at the A6 tunnel in the YangYang laboratory was used as one input spectra. In addition, the combination of the Maxwellian distribution for thermal neutrons, 1/E spectrum for the epi-thermal region (re-scattering neutrons), and fission spectrum for fast neutrons was used as input spectrum with various ratios of the three components. The unfolding results with the different input spectra show a similar spectral shape. The average spectrum was considered as our final spectrum.

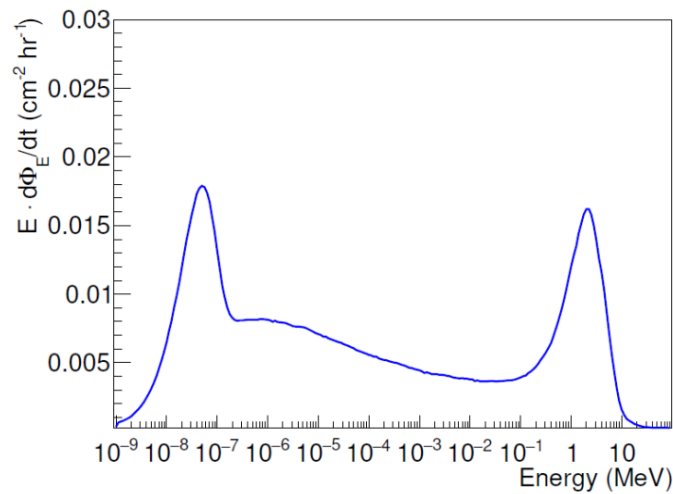

Figure 8: Neutron energy spectrum at A5 laboratory. Averaged output of unfolding processes using various input spectra was considered as a neutron spectrum.

## 4. Results and Discussion

### 4.1. Neutron energy spectrum

Figure 8 shows the neutron energy spectrum in the A5 laboratory obtained from the measurement and unfolding process. The vertical axis of Fig. 8 is the flux density per logarithmic energy bin $\frac{d\Phi_{logE}}{dt}$, or the neutron energy times the differential flux $E \cdot \frac{d\Phi_E}{dt}$, where the two expressions are same because $\Phi_{logE} = \frac{d\Phi}{d(logE)} = E \cdot \frac{d\Phi}{dE} = E \cdot \Phi_E$. The horizontal axis is E in MeV with a logarithmic scale.

The red dot spectrum with uncertainties at figure 9 shows the re-binned spectrum of figure 8 with wider bins. With the wider bins, the reasonably large statistics can be obtained to estimate the statistical uncertainties.

The neutron flux in the A5 laboratory can be obtained from the neutron spectrum. The neutron flux is $(4.46 \pm 0.66) \times 10^{-5}$ cm$^{-2}$ s$^{-1}$ in total. The thermal neutron flux is $(1.44 \pm 0.15) \times 10^{-5}$ cm$^{-2}$ s$^{-1}$ and the fast neutron flux (1 - 10 MeV) is $(0.71 \pm 0.10) \times 10^{-5}$ cm$^{-2}$ s$^{-1}$. The proportion of the thermal and fast neutron flux to the total flux is 32% and 16%, respectively. The uncertainties of the fluxes are the combination of the uncertainty of the BS count and that of the estimation of input spectra which is assumed to be approximately 10%.

The measured neutron energy spectrum shows a clear high-energy peak (fast neutron peak) in the region between 0.1 MeV and 10 MeV and a clear thermal peak. The medium energy regions between 1 eV and 0.1 MeV show relatively lower values in the vertical axis. The fast neutron peak shows that the neutrons produced by the ($\alpha$, n) reaction do not lose their energy much during transport to the point of measurement. The thermal neutrons or medium energy neutrons are produced through the kinetic energy loss of fast neutrons via the successive scatterings with nuclei inside materials such as rocks, air, or experimental structures.

### 4.2. Comparison with neutron energy spectrum at A6 laboratory, YangYang

The neutron measurement at the A5 laboratory was a repeat of the measurement made at the A6 laboratory using the BSS system. The neutron measurements at the A6 laboratory in 2010 covered the energy range up to the GeV level, using additional four metal-shell BSs, while the measurements at the A5 laboratory covered the range up to 20 MeV. The same unfolding process was used to estimate fluence rates for measurement at both laboratories. The input spectrum for the unfolding processes were mostly the same and one of the input spectra used for measurement at the A5 laboratory was a neutron spectrum measured at the A6 laboratory.

Figure 9 shows the comparison of the current measurements at the A5 laboratory and the 2010 measurements made at the A6 laboratory [3]. The total neutron flux in the two laboratories was similar, $(67.2 \pm 2.2) \times 10^{-6}$ $cm^{-2}s^{-1}$ at A6. However, the energy spectrum at the A6 laboratory is quite different from that of the A5 laboratory. The neutron energy spectrum at the A6 laboratory does not show the clear peak for fast neutrons (above 1 MeV) and shows an increase (and more neutrons) in the medium energy region and more thermal neutrons. From the spectrum, it can be assumed that the fast neutrons produced at the A6 laboratory have experienced more scattering or lost more energy per scattering than those in the A5 laboratory. In addition, the rocks or air in the A6 tunnel may contain more hydrogen nuclei, most likely in the form of water or gas. The total neutron flux measurement shows that the production of neutrons through ($\alpha$, n) reactions is similar for both cases.

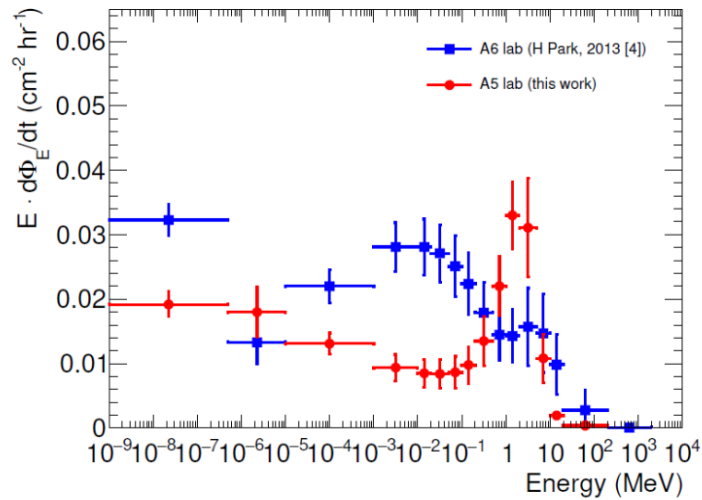

Figure 9: The neutron energy spectrum measured at the A5 laboratory (circles) and the A6 laboratory (squares) in YangYang Underground Laboratory.

The neutrons produced travel through the rock before they reach the detector. The energy of the neutrons at the location of the detector depends on the number of scatterings and the mass number of the target nuclei. The A6 and A5 laboratories are several hundred meters away from each other, but in the same mountain. However, the humidity in the cavity and the fraction of water ($H_2O$ molecule) in the rock may be different. The humidity in the cavity and the water containment in the rock vary according to the seasons and are affected by intermittent heavy rain. High humidity and/or more water in the rock in the area of the A6 laboratory could cause a softer neutron energy spectrum than that in the A5 laboratory. The total neutron fluxes of the two measurements are similar in spite of the difference in spectral shape. This suggests that the neutron production itself is similar for both areas.

## 5. Comparison of underground measurements

Many measurements for the neutron flux in underground laboratories have been performed and the results have been reported. Figure 10 shows the total, thermal, and fast neutron fluxes at various underground laboratories. The neutron fluxes are not correlated to the vertical overburden at the laboratory, as expected.

The neutron energy spectrum produced in the rock should follow the neutron energy spectrum of ($\alpha$, n) reactions, i.e. mainly distributed from 0.1 to 20 MeV. During travel through the rock, air, and structures inside laboratories, the neutrons slow down due to successive scattering. Some neutrons could disappear because of neutron capture or nuclear reaction. However, the nuclear reaction cross section or the neutron capture cross section for nominal nuclear elements inside rock is much smaller than the elastic scattering cross section. Unless there is no significant water storage inside the rock, the total neutron flux should be much more stable than the thermal or fast neutron flux. Therefore, the total neutron flux is the best choice when comparing the measures of the neutron fluxes in various laboratories.

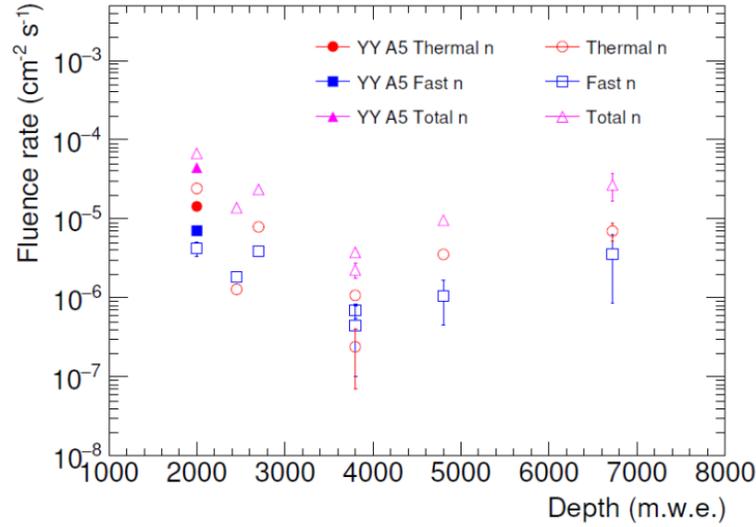

Figure 10: Total, thermal, and fast neutron fluxes at the YangYang Underground Laboratory (2000 m) (filled triangles, circles, and squares, respectively) compared to other laboratories (open triangles, circles, and squares, respectively) listed in Table 2.

The total, thermal, and fast neutron fluxes in underground laboratories are shown in Table 2. The total neutron flux in underground laboratories in Europe, such as Canfranc, Gran Sasso National Laboratory (LNGS), and Modane, are between $2.26 \times 10^{-6}$ and $13.8 \times 10^{-6}$ cm$^{-2}$ s$^{-1}$, whereas rates in laboratories in East Asia such as YangYang, Kamioka, and China Jinping Underground Laboratory (CJPL) are between $23.5 \times 10^{-6}$ and $67.2 \times 10^{-6}$ cm$^{-2}$ s$^{-1}$.

The total neutron flux is different at each site most probably due to the different neutron production rate by the ($\alpha$, n) reactions at each site. The ($\alpha$, n) reaction rate depends on the number of $\alpha$ particle sources ($^{238}$U and $^{232}$Th) and also the amount of specific target material having a significant ($\alpha$, n) reaction cross section inside the rock.

Table 2: Total, thermal, and fast neutron fluxes (× 10⁻⁶ cm⁻² s⁻¹) in underground laboratories

| Underground laboratory | Depth (m.w.e.) | Thermal neutron | Fast neutron [1 - 10 MeV][a] | Total neutron | Detector type |
|---|---|---|---|---|---|
| YangYang A6 [3] | 2000 | 24.2 ± 1.8 | 4.2 ± 0.9 | 67.2 ± 2.2 | ³He PC BS[b] |
| YangYang A5 | 2000 | 14.4 ± 1.5 | 7.1 ± 1.0 | 44.6 ± 6.6 | ³He PC BS |
| Canfranc [24] | 2450 | 1.28 ± 0.04 | 1.84 ± 0.03 | 13.8 ± 1.4 | ³He PC PE[c] |
| Kamioka [13] | 2700 | 7.88 | 3.88 | $23.5 \pm 0.7^{+1.9}_{-2.1}$ | ³He PC[c] |
| LNGS Hall A [11, 25] | 3800 | 1.08 ± 0.02 | 0.7 ± 0.14 | 3.78 ± 0.25 | BF₃ PC PE |
| LNGS Hall C [25] | 3800 | 0.24 ± 0.17 | 0.45 ± 0.35 | 2.26 ± 0.49 | LS[d] and MC |
| Modane [26] | 4800 | 3.57±0.05±0.27 | 1.06 ± 0.1 ± 0.6 [ > 1 MeV ] | 9.6 | ³He PC PE |
| CJPL-I [14] | 6720 | 7.03 ± 1.81 | 3.63 ± 2.77 [ 1-20 MeV] | 26.9 ± 10.2 | ³He & ⁴He PC BS |

[a] Energy range for fast neutron measurements.
[b] proportional counters with Bonner spheres
[c] proportional counters with Polyethylene
[d] Liquid scintillator

Table 3 shows the concentration of ²³⁸U and ²³²Th in the rock samples for each underground laboratory. In YangYang, the U/Th concentrations were measured by inductively coupled plasma mass spectrometry and found to be 3.9 ± 1.4 ppm (²³⁸U) and 10.5 ± 6.5 ppm (²³²Th) for the A5 laboratory and 2.1 ± 0.4 ppm (²³⁸U) and 13.0 ± 1.2 ppm (²³²Th) for the A6 laboratory [27]. From Tables 2 and 3, the concentration of U/Th does not reflect a good correlation to the total neutron flux. The concentrations of the ²³⁸U and ²³²Th in the YangYang Underground Laboratory are 5 to 10 times higher than those in the other laboratories, but this is not reflected in the total neutron flux (especially for Kamioka and CJPL). Also, LNGS and Modane laboratories have similar concentrations of U/Th but these are not reflected in the total neutron flux.

Table 3: Concentration of $^{238}$U and $^{232}$Th in the rock

| Underground laboratory | $^{238}$U (ppm) | $^{232}$Th (ppm) |
|---|---|---|
| YangYang A6 [27] | 2.1 ± 0.4 | 13.0 ± 1.2 |
| YangYang A5 | 3.9 ± 1.4 | 10.5 ± 6.5 |
| Kamioka [13] | 0.6 | 1.3 |
| LNGS Hall A[25] | 6.80 ± 0.67 | 2.167 ± 0.074 |
| LNGS Hall C[25] | 0.66 ± 0.14 | 0.066 ± 0.025 |
| Modane [9] | 0.84 ± 0.2 | 2.45 ± 0.2 |
| CJPL-I [28] | 0.30-0.34 | 0.13-0.16 |

To explain the differences in the neutron flux, the target nuclei of ($\alpha$, n) reactions need to be compared. The chemical composition of the rocks in the underground laboratories are shown in Table 4. The chemical composition values of YangYang are from rock samples and outcrop samples in the YangYang area [29], and not from the underground laboratories themselves. The composition of the YangYang and Kamioka igneous rocks are similar because both are volcanic. The rocks from Kamioka sample 1, LNGS, and Modane contain more Calcium (Ca). The YangYang samples and both Kamioka samples have more Silicon (Si).

Figure 11 shows the cross section of neutron production for ($\alpha$, n) and ($\alpha$, np) reactions with various nuclei which are the main components of the rocks in the underground laboratories. For an element, the cross section of neutron production for each isotope was averaged with its natural abundance. All $X_{nat}$($\alpha$, 2n) and $X_{nat}$($\alpha$, 3n) reactions require $\alpha$ particles with energy over 10 MeV. The cross section data are sourced from the Experimental Nuclear Reaction database in the National Nuclear Data Center (NNDC) site [33, 34]. The light nuclei such as Li, Be, and B, which have large ($\alpha$, n) cross sections, are also shown. The small concentration of those nuclei in the rock can make a significant contribution to the neutron flux. All $\alpha$ particles from the $^{238}$U and $^{232}$Th decay chains have energy ranging from 4 to 9 MeV and mainly in the range from 4 to 6 MeV. These results suggest that Sodium (Na) and Ca could be the main source of the neutron production inside rocks. The rocks in YangYang and the Kamioka igneous rock contain significant amounts of Na (about 3% of the total). Ca is the main component (about 30 %) for Modane, LNGS, and Kamioka sample 1. Variations in the amount of Na or Ca in the rock can result in a different neutron flux for each underground laboratory. In addition, it has been reported that the number of B and Be nuclei contained in the rocks, even in rocks of the same type, can vary by a factor of ten to a hundred [30, 31, 32]. The concentration of B and Be in the rock can vary the total neutron rate significantly, as well.

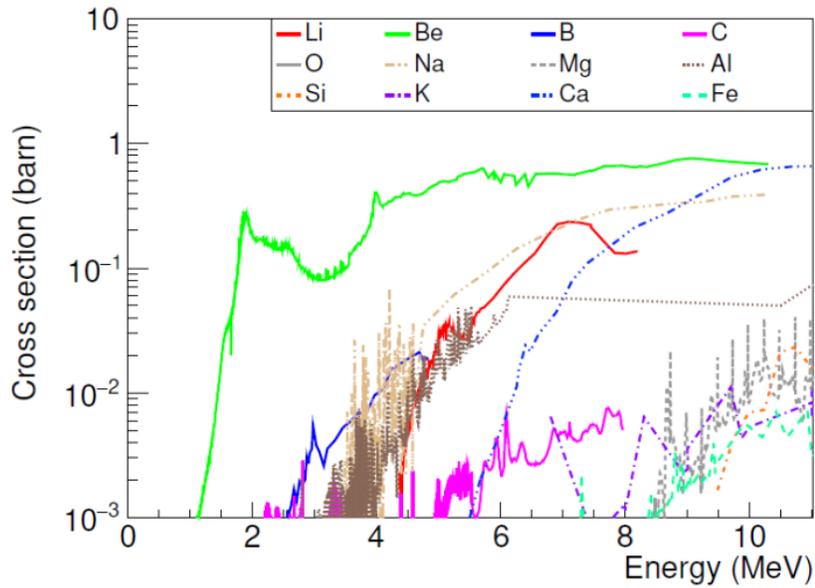

Figure 11: Cross sections of neutron production $X_{nat}(\alpha, n)$ and $X_{nat}(\alpha, np)$ reactions for the various elements.

Table 4: Concentration of chemical compositions in the rock or soil (% weight) for the various underground laboratories. The details are shown in the text.

| Underground laboratory | C | O | Na | Mg | Al | Si | K | Ca | Fe |
|---|---|---|---|---|---|---|---|---|---|
| YangYang [a] [29] | - | 48 | 2.6 | 2.4 | 8.4 | 26 | 1.9 | 4.2 | 5.2 |
| Kamioka sample 1 [13] | - | 39 | 0.01 | 0.6 | 6.0 | 17 | - | 28 | 7.6 |
| Kamioka igneous rock[b] [13] | - | 47 | 2.7 | 1.2 | 7.5 | 32 | - | 2.5 | 2.6 |
| LNGS [25] | 11.88 | 47.91 | - | 5.58 | 1.03 | 1.27 | 1.0 | 30.29 | - |
| Modane[c] [9] | - | 24 | - | 5.6 | 1.0 | 1.3 | 0.1 | 30 | - |

[a] Average of twelve samples from YangYang area, 6 drill hole and 6 outcrop samples
[b] Average of two igneous rock samples, which are widely distributed around Kamioka area
[c] Weight loss on ignition ($H_2O$, $CO_2$, etc) was 31.5%.

## 6. Summary

The neutron flux in the YangYang A5 laboratory was measured with a BSS system using $^3$He proportional counters. The total neutron flux measured was $(4.46 \pm 0.66) \times 10^{-5}$ cm$^{-2}$ s$^{-1}$. The thermal neutron flux was $(1.44 \pm 0.15) \times 10^{-5}$ cm$^{-2}$ s$^{-1}$ and the fast neutron flux, ranging from 1 MeV to 10 MeV, was $(0.71 \pm 0.10) \times 10^{-5}$ cm$^{-2}$ s$^{-1}$. The total neutron flux in the A5 laboratory was approximately 40% lower than that in the A6 laboratory (measured in 2010). The neutron energy spectrum in the A5 laboratory is different compared to that in the A6 laboratory. The neutron flux at the A5 laboratory is composed of more fast neutrons and fewer medium energy (epi-thermal) neutrons (1 eV – 0.1 MeV). The difference in the energy spectrum can be explained by the different neutron transports in the path from the generation to the point of measurement such as the water content in the rock or air.

Currently, in the YangYang A5 laboratory, two experiments, COSINE and AMoRE, are being conducted and being upgraded to the next phase. This study will be used in the background modeling analysis as well as in the design of shielding systems for both these experiments.

**Acknowledgement**

This research is supported by the Korea Research Institute of Standards and Science Grant No. 20011032. We acknowledge the support received from the Center for Underground Physics, the Institute for Basic Science in Korea and the Korea Hydro and Nuclear Power Company.